\documentclass[a4paper,fleqn,useAMS]{mnras}

\usepackage{graphicx}%

\title[Local Volume Late-Type Galaxies with an Extreme Ratio of $H\alpha$-to-$FUV$ Star-Formation Rates]
{\bf{}Local Volume Late-Type Galaxies with an Extreme Ratio of $H\alpha$-to-$FUV$ Star-Formation Rates}

\author[I.\,D.\,Karachentsev, E.\,I.\, Kaisina, S.\,S.\,Kaisin]
{I.\,D.\,Karachentsev$^{1}$\thanks{E-mail:ikar@sao.ru},
{E.\,I.\,Kaisina$^{1}$},
{S.\,S.\,Kaisin}$^{1}$\\
$^{1}$Special Astrophysical Observatory of the Russian Academy
 of Sciences, Nizhnij Arkhyz, KChR, 369167, Russia}

\begin{document}
\maketitle
\begin{abstract} Using a currently most representative sample of 477 late-type galaxies within 11 Mpc of the Milky Way with measured star-formation rates ($SFR$s) from the far ultraviolet ($FUV$) and H$\alpha$ emission line fluxes, we select galaxies with the extreme ratios: $SFR(H\alpha)/SFR(FUV) > 2$ and $SFR(H\alpha)/SFR(FUV) < 1/20$. Each subsample amounts to $\sim5$\% of the total number and consists of dwarf galaxies with the stellar masses $M^*/M_{\odot} = (5.5 - 9.5)$~dex. In spite of a huge difference in their $SFR(H\alpha)$ activity on a scale of $\sim10$~ Myr, the temporarily ``excited'' and temporarily ``quiescent'' galaxies follow one and the same relation between $SFR(FUV)$ and $M^*$ on a scale of $\sim100$~Myr. Their average specific star-formation rate $\log[SFR(FUV)/M^*] = -10.1\pm0.1$ (yr$^{-1}$) coinsides with the Hubble parameter $\log(H_0)= -10.14$ (yr$^{-1}$). On a scale of $t \sim10$~Myr, variations of $SFR$ have a moderate flash amplitude of less than 1 order above the main-sequence and a fading amplitude to 2 orders below the average level. In general, both temporarily excited and temporarily quiescent galaxies have almost similar gas fractions as normal main-sequence galaxies, being able to maintain the current rate of star-formation on another Hubble time scale. Ranging the galaxies according to the density contrast produced by the nearest massive neighbor exhibits only a low average excess of $SFR$ caused by tidal interactions. 

\end{abstract} 


\section{Introduction} Big disc-shaped galaxies with stellar masses of $M^*/M_{\odot}=(9.5-11.0)$~dex have a current star-formation rate ($SFR$) approximately proportional to the galaxy stellar mass. For spiral galaxies without apparent bulges, the specific star-formation rate, $sSFR=SFR/M^*$, is close to the Hubble parameter $H_0=-10.14$~dex (yr$^{-1}$), smoothly dropping at $\sim0.5$~dex towards the most massive discs (Bothwell et al. 2009;
Schiminovich et al. 2010; Karachentsev et al. 2013a; Karachentsev \& Kaisina, 2019). Based on Karachentseva et al. (2020), the cosmic variation of bulgeless galaxies according to the regression line $sSFR(M^*)$, named the ``main sequence'' (Sparre et al. 2015), is equal to 0.16~dex only. This means that the discs of spiral galaxies uniformly convert their gas into stars at a rate roughly the same on the whole cosmological scale $H_0^{-1}$. Late-type Sm and Im galaxies of medium masses also follow the ``main sequence'' with the parameter $sSFR\simeq H_0$ (James et al. 2008). Such a quiescent process can be called ``smouldering'' star formation. 

Histories of star formation in dwarf galaxies with the masses $M/M_{\odot}\sim(5-9)$~dex are more varied. This is mainly due to the shallow potential well of dwarf galaxies, where the gaseous component is easily subject to both external and internal perturbations. The most suitable sample for studying the star-formation history in dwarf galaxies is the Local Volume (LV) comprising about 1000 galaxies with the distances $D\leq 11$~Mpc (Karachentsev et al. 2013b = Updated Nearby Galaxy Catalog, UNGC). Dwarf galaxies with the stellar masses $M^*/M_{\odot}<9.5$~dex make the predominant population $(\sim5/6)$ of this sample. 

Most LV galaxies have measurements of the integral flux in the far ultraviolet ($FUV, \lambda_{eff}=1539$\AA, $FWHM=269$\AA) conducted with the GALEX space telescope (Martin et al. 2005) which allows one to estimate their star-formation rate on a characteristic scale of $\sim100$~Myr. The systematic review of LV galaxies in the H$\alpha$ emission line was carried out by Kennicutt et al. (2008) and Kaisin \& Karachentsev (2019 and references therein). By now, the H$\alpha$ fluxes are measured for more than 60\% of LV galaxies which yields the $SFR$ estimates for them on a time scale of $\sim10$ Myr. Comparison between $SFR(FUV)$ and $SFR(H\alpha)$ offers opportunities to detect galaxies, the star-formation rate of which is subject to great time variations. The observed data on the $SFR(H\alpha)/SFR(FUV)$ ratio for galaxies were discussed in the papers by Lee et al. (2009), McQuinn et al. (2009), Weisz et al. (2012), Karachentsev \& Kaisina (2013, 2019). On the whole, for massive galactic discs, the condition $SFR(H\alpha)\simeq SFR(FUV)$ works. With the decrease of luminosity (mass) of a late-type galaxy, the mean $SFR(H\alpha)/SFR(FUV)$ ratio 2--3 times drops, while the dispersion of the ratio increases considerably. Due to the feedback, the star formation shows the character of stochastic oscillations which is most pronounced in the most low-mass galaxies. Based on Stinson et al. (2007), Lee et al. (2009), Weisz et al. (2012), and Flores Velazques et al. (2021), the $SFR$ variations can reach an amplitude of $\sim30$ at the time interval of $\sim30$~Myr with a typical quasi-period of $\sim300$~Myr. In addition to the $SFR$ variability due to inner causes, the star-formation rate of dwarf galaxies located near massive neighbors undergoes quenching because of sweeping-out of gas, when they move through the neighbor halo. Irregular dwarfs turn into spheroidal (Irr$\rightarrow$ Tr $\rightarrow$ Sph) losing the conditions for further star formation. According to calculations by Akins et al. (2021), dwarf galaxies with the active star formation and stellar masses $M^*/M_{\odot}\sim(6-8)$~dex transform into quiescent systems over $\sim2$~Gyr. Therein, more massive dwarfs with $M^*/M_{\odot}>8$~dex appear almost insensitive to external quenching. 
 From another point of view, tidal disturbances from a tight massive neighbor can lead to an increase in the rate of star formation in a gas-rich dwarf
galaxy. In general, the study of galaxies with extreme values of the ratio $SFR(H\alpha)/SFR(FUV)$ can shed light on the features of the process of star formation in late-type galaxies on a time scale of 10-100 Myr.

\section{Star-formation rate recipes}

The updated version of the UNGC catalog is presented in the database: http://www.sao.ru/lv/lvgdb. It includes 477 late-type galaxies ($T>4$), i.e., of the  spiral types Sc, Sd, Sm, blue compact dwarfs (BCD), irregular types Im, Irr, and of the transient type (Tr) with $SFR$ measured both from the $FUV$ flux and via the H$\alpha$ flux. The data on them are collected in Appendix A. Brief description of some individual cases is given in Appendix B. We use these data to search for galaxies with significant divergence of the $SFR(H\alpha)$ and $SFR(FUV)$ values. Analysis of such aberrant cases is the main subject of our paper. 

According to Kennicutt (1998) and Lee et al. (2009), the integral star-formation rate of a galaxy in units $(M_{\odot}/$yr) is expressed as 
\begin{equation}
 \log[SFR(FUV)]=2.78+2\log D-0.4\times m^c_{FUV}, 
\end{equation}
where $D$ is the distance in Mpc and $m^c_{FUV}$ is the apparent magnitude of a galaxy in the $FUV$ band corrected for the Galactic and internal extinction following a recipe of Lee et al. (2009):
 \begin{equation}
   m^c_{FUV} = m_{FUV} - 1.93( A^G_B + A^i_B).
 \end{equation}
 Here, the Galactic extinction in the B-band, $A^G_B$, was taken from Schlegel et al. (1998), and the internal extinction in the galaxy itself was determined as
\begin{equation}
A^i_B = [1.54 + 2.54 (\log 2 V_m - 2.5)] \log (a/b) 
\end{equation}
through the apparent galaxy axis ratio $a/b$ and the amplitude of internal rotation $V_m$ (Verheijen, 2001). For dwarf galaxies with $ V_m < 39 km s^{-1}$ the internal extinction was considered negligible. The data on ratios $a/b$ and rotation amplitudes $V_m$ were taken from UNGC (see details on the method to compute $V_m$ in Karachentsev et al. 2013b).
  
The integral star-formation rate of a galaxy $(M_{\odot}/$yr) determined by its H$\alpha$ flux, $F_c(H\alpha)$, is written as 
\begin{equation}
\log[SFR(H\alpha)]=8.98+2\log D+ \log F_c(H\alpha),
       \end{equation}
where the $H\alpha$ flux in (erg$\times$ cm$^{-2}$ sec$^{-1}$) is corrected for extinction according to Lee et al. (2009):
\begin{equation}
 \log F_c(H\alpha) = \log F(H\alpha) + 0.215( A^G_B + A^i_B).
\end{equation}
\begin{figure}
 \includegraphics[height=0.35\textwidth]{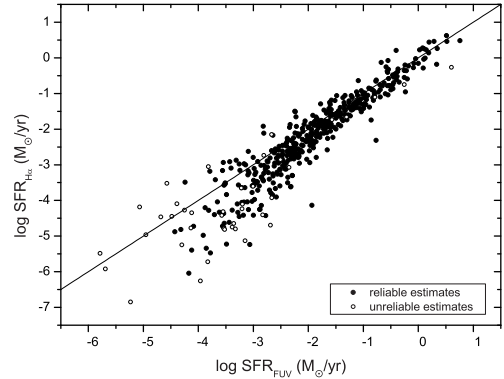}
\caption{Distribution of the Local Volume galaxies of the $T>4$ morphological types on the integral star-formation rate measured from the $H\alpha$ and $FUV$ fluxes. The open circles denote the galaxies with unreliable estimates of the $H\alpha$ or $FUV$ fluxes. The diagonal line corresponds to the equation $SFR(H\alpha)=SFR(FUV)$.} \end{figure} 

Figure 1 shows the distribution of LV galaxies with $T>4$ on two independent estimates of $\log(SFR)$. The open circles refer to the cases, when the galaxies have only the upper limit of $SFR(H\alpha)$ or $SFR(FUV)$ measured. This symbol also indicates several galaxies with unreliable $SFR(H\alpha)$ estimates due to observations in varying weather conditions and also the large spiral galaxy IC 342, the H$\alpha$ flux of which is measured only in the central part of its disc . As one can see, the range of $SFR$ estimates spans six orders from $\sim10$ to $\sim10^{-5} M_{\odot}$/yr. Typical measurement accuracy of the H$\alpha$ and $FUV$ fluxes for the Local Volume galaxies is about 15\%  (Kaisin et al. 2011). But, with $SFR<10^{-4}M_{\odot}$/yr, measurement errors become prevailing. In the region of the reliable $SFR$ estimates, the scatter of galaxies relative to the diagonal line $SFR(H\alpha)=SFR(FUV)$ significantly exceeds measurement errors and is due mainly to the difference in the activity of galaxies at times of $\sim10$~Myr and $\sim100$~Myr.
 However, differences in stellar initial mass function and metallicity may also be reponsible for the discrepancy between $SFR(H\alpha)$ and $SFR(FUV)$
(Lee et al. 2009 and references therein).
\begin{figure*}
 \includegraphics[height=0.35\textwidth]{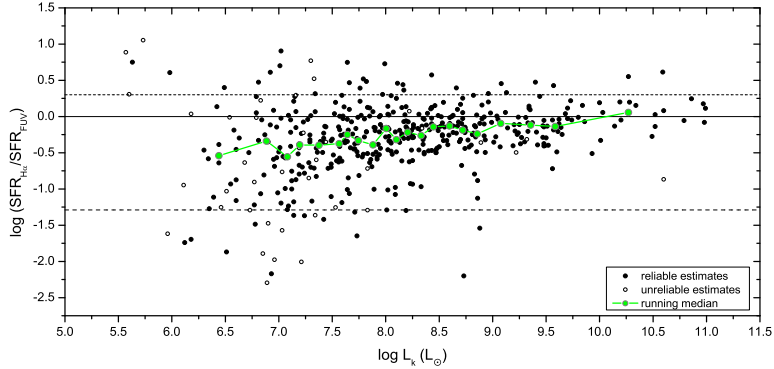}
\caption{Ratio of star-formation rates measured from the H$\alpha$ and $FUV$ fluxes for the LV galaxies having different integral luminosities in the K band. Indications of galaxies are the same as in Fig.1. Two dashed lines mark the regions of temporarily excited and temporarily quiescent galaxies. The broken green line indicates the running median} \end{figure*} 

Figure 2 represents the distribution of late-type LV galaxies on the $SFR(H\alpha)/SFR(FUV)$ ratio corrected for extinction and the integral luminosity in the K band. Indications of galaxies with different symbols are the same as in Fig.1. The broken green line shows the running median with a window of 20 objects having reliable flux estimates.

 We excluded from consideration the bright spiral galaxy IC~342 with the underestimated H$\alpha$ flux from its periphery and also two bright spiral galaxies: M~82 ($L_K=10.59$~dex) and NGC~4517 ($L_K=10.27$~dex) with the $\log[SFR(H\alpha)/SFR(FUV)$] ratios 0.61 and 0.57, respectively. These two galaxies stand out with the presence of strong shredded internal extinction and extraplanar H$\alpha$ filaments.  

The Fig.2 demonstrates two main features: the tendency of the mean $SFR(H\alpha)/SFR(FUV)$ ratio to decrease from massive galaxies to dwarf galaxies, and the increase of the dispersion of this ratio towards low-mass galaxies. Two horizontal dashed lines mark the region of the temporarily excited galaxies with $SFR(H\alpha)/SFR(FUV) >2$ and the region of the temporarily quiescent galaxies with $SFR(H\alpha)/SFR(FUV) <1/20$. The indicated limits are chosen in order, on the one hand, to extinguish the effect of flux measurements errors, and on the other hand, to have sufficiently representative subsamples of galaxies.

\begin{figure}
 \includegraphics[height=0.35\textwidth]{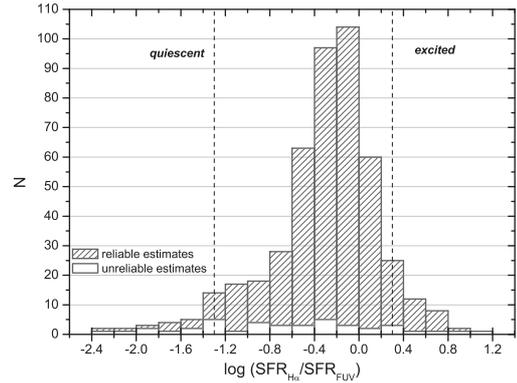}
 \caption{Distribution of the number of LV galaxies with $T>4$ on the $SFR(H\alpha)/SFR(FUV)$ ratio. The galaxies with unreliable H$\alpha$ or $FUV$ fluxes are not shaded.} \end{figure} 

The histogram in Fig.3 shows the distribution of the number of galaxies on the $SFR(H\alpha)/SFR(FUV)$ ratio. The unshaded area of the histogram along the horizontal axis corresponds to the galaxies with unreliable fluxes in H$\alpha$ or $FUV$. The objects at the right and left tails of this distribution are the subject of our further consideration. 

\section{Temporarily excited and temporarily quiescent dwarfs} 

\begin{table*}
\caption{Late-type ($T>4$) LV galaxies with $SFR(H\alpha)/SFR(FUV) > 2$}
\begin{tabular}{llrrccrrrr} \hline

  Name      &   $T$  &  $D$   &$\Theta_1$ &$\log(L_K)$ &$\log(M_{HI})$ &$\log SFR(H\alpha)$& $\log SFR(FUV)$& $\Delta \log(SFR)$& Ref.\\
\hline
            &      &  Mpc &       & $L_{\odot}$ & $M_{\odot}$ & $M_{\odot}$/yr & $M_{\odot}$/yr &   & \\
\hline
  (1)       &  (2) & (3)  &  (4)  &  (5)    &   (6)    &  (7)     &   (8)      &   (9) & (10)\\ 
\hline
PGC 2448110 &  Irr & 6.98 &  5.2  &  7.02   &    $-$     & $-$1.92    &  $-$2.82     &  0.90 &  1
 \\
Clump III   &  Irr & 3.60 &  3.1  &  5.57   &    $-$     & $-$4.18    & $<-$5.07     & $>$0.89 &  2
 \\
KKH 6       &  Irr & 5.18 & $-$0.1  &  7.30   &   6.92   & $-$3.05    & $<-$3.82     & $>$0.77 &  3
 \\
JKB 83      &  Irr & 3.70 &  5.3  &  5.63   &    $-$     & $-$3.49    &  $-$4.24     &  0.75 &  1
\\
Mrk 475     &  BCD & 9.20 & $-$1.2  &  7.64   &   6.48   & $-$1.50    &  $-$2.24     &  0.74 &  4
\\
UGC 8105    &  Im  & 9.99 & $-$0.3  &  7.99   &   7.82   & $-$1.51    &  $-$2.24     &  0.73 & 12 \\
JKB 142     &  Irr &10.50 &  0.5  &  7.01   &   7.65   & $-$2.11    &  $-$2.81     &  0.70 &  5
 \\
N2903-HI    &  Irr & 8.90 &  3.3  &  6.92   &   6.42   & $-$2.93    &  $-$3.54     &  0.61 &  6
\\
KDG 61em    &  Irr & 3.70 &  4.5  &  5.98   &    $-$     & $-$3.18    &  $-$3.79     &  0.61 &  7
\\
UGCA 298    &  BCD &11.00 & $-$1.1  &  8.43   &   7.29   & $-$1.65    &  $-$2.23     &  0.58 &  1\\  
Cam A       &  Irr & 3.56 &  0.7  &  7.79   &   8.14   & $-$2.91    &  $-$3.43     &  0.52 &  3
\\
LV J1157+56 &  Irr & 8.75 & $-$0.8  &  7.33   &    $-$     &$>-$2.15    &  $-$2.67     & $>$0.52 &  9
\\
UGCA 281    &  Irr & 5.70 & $-$0.9  &  7.82   &   7.77   & $-$1.29    &  $-$1.78     &  0.49 & 10
\\
NGC 1569    &  Sm  & 3.19 &  1.1  &  9.40   &   8.30   & $-$0.08    &  $-$0.55     &  0.47 &  3
\\
KDG 10      &  Irr & 7.87 & $-$1.5  &  7.66   &   7.72   & $-$2.62    &  $-$3.09     &  0.47 & 11
\\
Garland     &  Irr & 3.82 &  3.0  &  6.81   &   7.54   & $-$2.17    &  $-$2.64     &  0.47 &  7
\\
KK 65       &  Irr & 7.98 & $-$0.1  &  8.11   &   7.70   & $-$1.87    &  $-$2.33     &  0.46 &  3
\\
NGC 5253    &  Sm  & 3.44 &  0.3  &  9.08   &   7.91   & $-$0.64    &  $-$1.09     &  0.45 &  8
\\
IC 3077     &  BCD & 9.10 &  0.1  &  8.16   &   6.97   & $-$2.87    &  $-$3.31     &  0.44 & 14\\ 
NGC 3077    &  Im  & 3.85 &  2.3  &  9.57   &   8.80   & $-$1.06    &  $-$1.48     &  0.42 &  7
\\                  
NGC2683dw1  &  Irr & 9.82 &  4.3  &  6.49   &    $-$     & $-$3.16    &  $-$3.56     &  0.40 &  9
\\
Mrk 1329    &  BCD & 9.80 & $-$0.4  &  8.66   &   8.24   & $-$1.19    &  $-$1.58     &  0.39 & 12 \\
ESO 553-46  &  BCD & 6.70 & $-$2.1  &  8.17   &   7.46   & $-$1.25    &  $-$1.61     &  0.36 & 13
\\
NGC 4861    &  Im  & 9.95 & $-$0.6  &  9.13   &   8.86   & $-$0.25    &  $-$0.59     &  0.34 & 12
\\
NGC 625     &  Sm  & 4.02 & $-$0.3  &  8.96   &   8.00   & $-$1.20    &  $-$1.52     &  0.32 &  8
\\
LV J0300+25 &  Irr & 7.80 &  1.6  &  7.34   &   6.20   & $-$3.09    &  $-$3.41     &  0.32 &  3
\\
KK 35       &  Irr & 3.16 &  2.4  &  7.97   &    $-$     & $-$1.61    &  $-$1.92     &  0.31 &  3
\\
NGC 4248    &  Sm  & 7.40 &  1.7  &  8.82   &   7.79   & $-$1.92    &  $-$2.22     &  0.30 &  8
\\
\hline
 & & & & & & & & & \\
 Mean       &   $-$  & 6.74 &  1.07 &  7.74   &   7.62   & $-$2.03    &  $-$2.56     &  0.53 &  $-
$\\
            &     &$\pm$0.48 &  0.38 &  0.20   &   0.16   &  0.20    &   0.20     &  0.03 & 
  \\
\hline
 
           & & & & & & & & & \\
\multicolumn{10}{l}{{\bf References:} 1 --- Kaisin+2019, 2 --- Karachentsev+2011, 3 --- Karachentsev+2010, 4 --- Kaisin+2014, 5 --- James+2017,} \\
\multicolumn{10}{l}{6 --- Kaisin+2013, 7 --- Karachentsev+2007, 8 --- Kennicutt+2008, 9 --- Karachentsev+2015a, 10 --- Kaisin+2008, }     \\
\multicolumn{10}{l}{11 --- Kaisin+2011, 12 --- Gil de Paz+2003,
13 --- Kaisin+2012, 14 --- Gavazzi+2006. }
 \end{tabular}
 \end{table*}

 \begin{table*}
 \caption{Late-type LV galaxies with  $SFR(H\alpha)/SFR(FUV) < 1/20$}
 \begin{tabular}{llrrccrrrr}\hline

  Name         &  $T$ &   $D$  & $\Theta_1$ & $\log (L_K)$ & $\log (M_{HI})$& $\log SFR(H\alpha)$ &$\log SFR(FUV)$ &$\Delta \log SFR$&  Ref.\\
 \hline
            &      &  Mpc &       & $L_{\odot}$ & $M_{\odot}$ & $M_{\odot}$/yr & $M_{\odot}$/yr &   & \\
\hline
   (1)       &  (2) & (3)  &  (4)  &  (5)    &   (6)    &  (7)     &   (8)      &   (9) & (10)\\ 
\hline
 ESO 410-005   & Tr &  1.93&   0.1  &  6.89  &  5.91  & $<-$6.26  & $-$3.96   &$<-$2.30  &  1
 \\
 DDO 120       & Im &  7.28&   1.1  &  8.73  &  8.04  & $ -$4.14  & $-$1.94   &$ -$2.20  &  2
 \\
 KDG 52        & Irr&  3.42&   2.0  &  6.93  &  7.02  & $ -$5.23  & $-$3.06   &$ -$2.17  &  3
\\
 HIPASSJ1337-39& Irr&  5.08&   0.0  &  7.21  &  7.60  & $<-$4.69  & $-$2.69   &$<-$2.00  &  9
\\
 KK 195        & Irr&  5.22&   0.9  &  6.96  &  7.56  & $<-$5.13  & $-$3.15   &$<-$1.98  &  1
\\
 ESO 540-030   & Tr &  3.56&   1.0  &  6.85  &  5.99  & $<-$5.72  & $-$3.83   &$<-$1.89  &  1
\\
 KKH 86        & Irr&  2.61&  $-$1.4  &  6.51  &  5.92  & $ -$6.04  & $-$4.17   &$ -$1.87  &  4
\\
 BK3N          & Irr&  4.17&   1.0  &  6.12  &   $-$    & $ -$5.23  & $-$3.49   &$ -$1.74  &  3
\\
 KK 230        & Irr&  2.21&  $-$1.3  &  6.18  &  6.35  & $ -$5.47  & $-$3.78   &$ -$1.69  &  5
\\
 LGS-3         & Tr &  0.65&   1.5  &  5.96  &  5.02  & $<-$6.85  & $-$5.23   &$<-$1.62  &  2
\\
 LSBC D564-08  & Irr&  8.83&  $-$0.2  &  7.03  &  7.54  & $<-$4.40  & $-$2.83   &$<-$1.57  &  2
\\
 N4656UV       & Irr&  7.98&   4.2  &  8.88  &  7.61  & $ -$2.31  & $-$0.77   &$ -$1.54  &  6
\\
 DDO 210       & Irr&  0.98&  $-$0.4  &  6.78  &  6.46  & $ -$5.34  & $-$3.86   &$ -$1.48  &  2
\\
 KDG 215       & Irr&  4.83&   0.5  &  6.90  &  7.38  & $<-$4.80  & $-$3.33   &$<-$1.47  &  7
\\
 UGCA 365      & Irr&  5.42&   0.7  &  7.73  &  7.29  & $ -$4.42  & $-$2.77   &$ -$1.45  &  8
\\
 KDG 56        & Irr&  8.90&   1.7  &  7.42  &  7.17  & $ -$4.42  & $-$3.00   &$ -$1.42  &  4
\\
 UGC 7298      & Irr&  4.19&  $-$0.4  &  7.24  &  7.27  & $ -$4.25  & $-$2.88   &$ -$1.37  &  5
\\
 KKH 80        & Irr&  6.14&  $-$1.2  &  7.14  &  6.92  & $ -$4.25  & $-$2.88   &$ -$1.37  &  5
\\
 LSBC D565-06  & Irr&  9.29&   0.5  &  7.34  &  6.76  & $<-$4.75  & $-$3.38   &$<-$1.37  &  2
\\
 UGC 5186      & Irr&  9.40&  $-$0.9  &  7.71  &  7.49  & $ -$3.90  & $-$2.58   &$ -$1.32  &  4
\\
 ESO 384-016   & BCD&  4.49&   0.4  &  7.83  &  6.69  & $<-$4.65  & $-$3.38   &$<-$1.29  &  2
\\
 HIDEEPJ1337-33& Irr&  4.55&   0.9  &  6.73  &  6.71  & $<-$4.81  & $-$3.52   &$<-$1.29  &  9
\\
 CGCG 014-054  & Im &  9.60&  $-$0.2  &  8.01  &  7.70  & $ -$3.61  & $-$2.32   &$ -$1.29  & 10\\  	
 \hline
  & & & & & & & & & \\
  Mean         & $-$  & 5.25 &  0.46  & 7.18   & 6.93   & $-$4.81   & $-$3.17    &$-$1.64   & \\
               &   &$\pm$0.56 &  0.27  & 0.15   & 0.16   &  0.21   & 0.19    & 0.07
  & \\
 \hline
 
 & & & & & & & & & \\
 \multicolumn{10}{l}{{\bf References:} 1 --- Bouchard+2009, 2 --- Kennicutt+2008, 3 --- Karachentsev+2007, 4 --- 
Kaisin+2011, 5 --- Kaisin+2008,}\\
 \multicolumn{10}{l}{6 --- Kaisin+2019, 7 --- Kaisin+2014, 8 --- Cote+2009,
 9 --- Grossi+2007, 10 --- LVGDB in preparation.}
 \end{tabular}
 \end{table*} 

Tables 1 and 2 present the basic observed data on the late-type LV galaxies with the extreme $SFR(H\alpha)/SFR(FUV)$ ratios. The first table includes 28 objects with $SFR(H\alpha)/SFR(FUV) \geq2$, the second contains 23 cases with $SFR(H\alpha)/SFR(FUV) \leq 1/20$. The table columns contain: (1) --- the name of a galaxy; (2) --- the morphological type; (3) --- the distance of a galaxy in Mpc; (4) --- the tidal index $\Theta_1$ described below in the section 5; it shows the density contrast created by the most significant neighbor, while the values $\Theta_1>0$ correspond to group members and $\Theta_1<0$ correspond to field galaxies; (5) --- the luminosity of a galaxy in the K band in $L_{\odot}$, corrected for extinction, (6) --- the hydrogen mass of a galaxy in $M_{\odot}$; (7, 8) --- the star-formation rate determined from the H$\alpha$ and $FUV$ fluxes in ($M_{\odot}$/yr); (9) --- the logarithm of the $SFR(H\alpha)/SFR(FUV)$ ratio; (10) --- the source of data on the flux in the H$\alpha$ line given in the table footnote. The data on the galaxies are taken from the database: http://www.sao.ru/lv/lvgdb (Kaisina et al. 2012) updated with new observations. The galaxies in the tables are ranged by the $SFR(H\alpha)/SFR(FUV)$ ratio. The last row in the tables shows the mean value of each parameter and error of the mean.

\begin{figure} 
\includegraphics[height=0.35\textwidth]{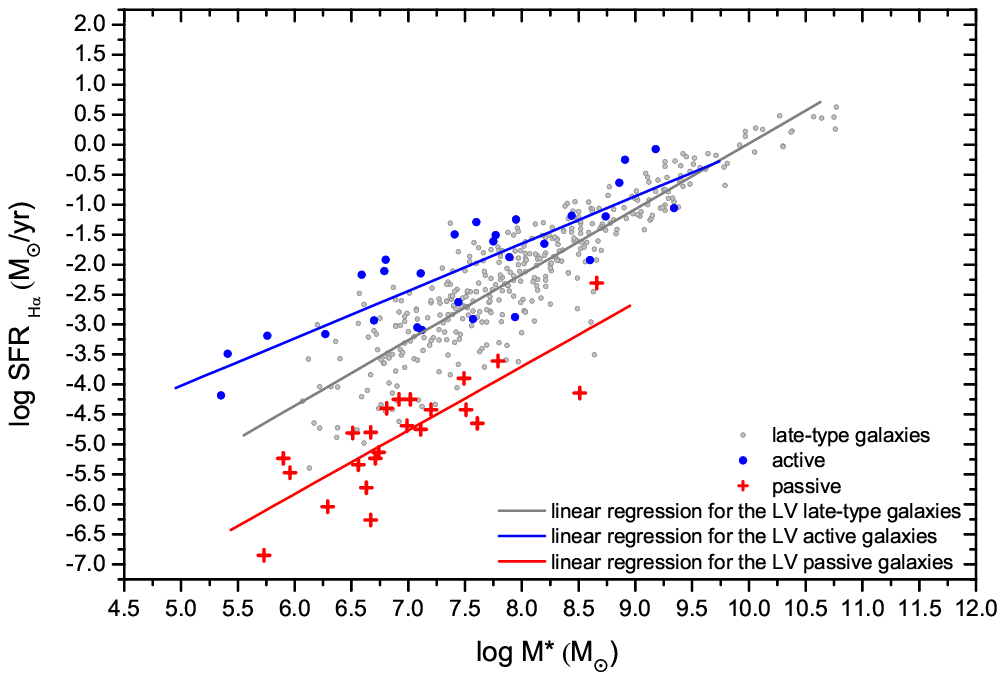}\\
\includegraphics[height=0.35\textwidth]{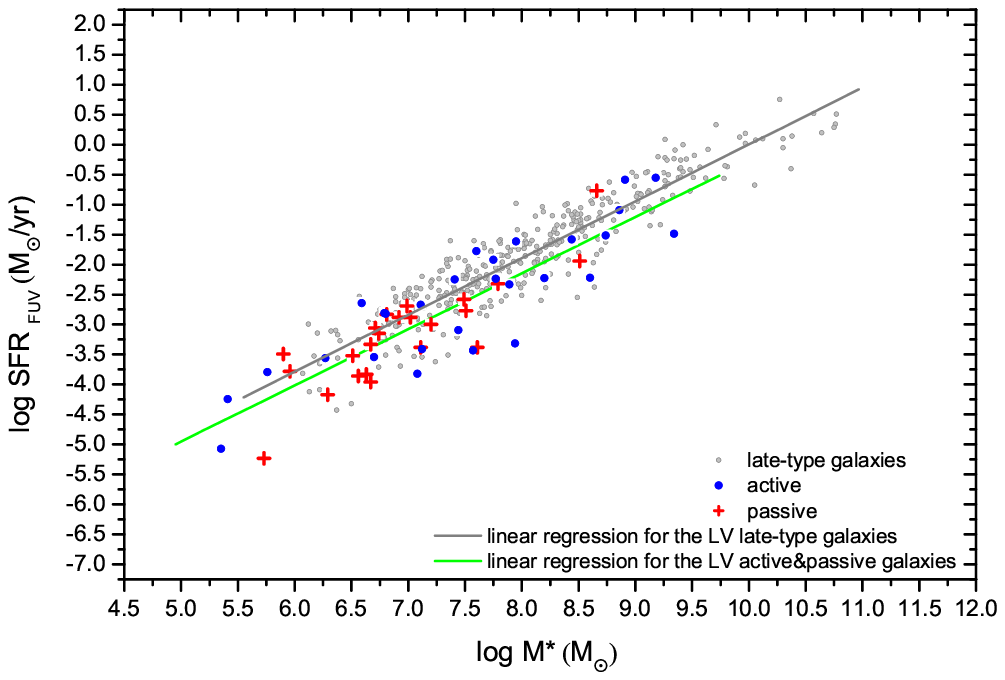}
\caption{Star-formation rate via the H$\alpha$ flux (the top panel) and via the $FUV$ flux (the bottom panel) for galaxies of different stellar masses
defined from eq. (5). Excited and quiescent galaxies are indicated by the solid circles and crosses, respectively. The main-sequence for other late-type galaxies from the whole parent sample is marked with gray dots. The parameters of linear regressions are specified in Table 3.} \end{figure} 

The top panel of Fig.4 shows the distribution of galaxies of both subsamples on $SFR(H\alpha)$ and the stellar mass $M^*$. Following Lelli et al. (2016), for conversion of $L_K$ to $M^*$, we accept the relation: 
\begin{equation}
 M^*/M_{\odot}=0.6\times L_K/L_{\odot}.
    \end{equation}
 Excited and quiescent galaxies are indicated by the solid circles and crosses, respectively. For comparison, other late-type galaxies from the whole our sample are shown as gray dots. Both subsamples of temporal divergents, as well as the whole sample galaxies, show positive correlations between $SFR$ and $M^*$, which are traced by corresponding regression lines. Three first rows in Table 3 give the linear regression coefficients for them, $Y=k\times X+c$. 

 As it is seen, the average amplitude of $SFR$ burst for the excited galaxies is characterized by a factor of (1 - 10) above the main-sequence, increasing towards the low mass objects, whereas the SFR for quiescent ones falls on average $\sim30$ times compared with the main-sequence.

The similar relation $SFR(FUV)$ vs $M^*$ is presented in the bottom panel of Fig.4. Unlike the data on upper panel, the excited and quiescent galaxies follow a single relationship described by the linear regression (the green line) 
\begin{equation}
\log[SFR(FUV)]=(0.94\pm0.07)\log M^*-9.6\pm0.5 
\end{equation}
with approximately the same spread. At total, the sequence of temporarily excited and temporarily quiescent dwarf galaxies roughly corresponds to the general sequence of late-type galaxies in the Local Volume, shown by gray dots. The regression line for them is indicated by the gray line, going 
$\sim0.25 dex$ above the green line. Their parameters are given in the fourth and the fifth rows of Table 3. Thus, a significant difference in SFRs between the excited and quiescent galaxies, visible on the $\sim10$ Myr time-scale via H$\alpha$ flux, become hardly distinguishable on the $\sim100$ Myr time-scale via $FUV$-flux.  

\section{Atomic hydrogen content and SFR bursts} Galaxies in the course of a starburst (N = 28) and in the course of quenching (N = 23) amount, respectively, 6\% and 5\% of the total number of late-type LV galaxies with the measured fluxes in H$\alpha$ and $FUV$. The majority of them (43 of 51) have hydrogen mass estimates derived from HI-flux as
\begin{equation}
\log(M_{HI})= 2.356 \times D^2 F_{HI},
\end{equation}
where $F_{HI}$ is in (Jy km s$^{-1}$) and $M_{HI}$ is in Solar masses. The main sources of data on HI-fluxes were sky surveys HIPASS (Koribalski et al. 2004), ALFALFA (Haynes et al. 2011), WSRT-CVn (Kovac et al. 2009) and a special survey of nearby dwarf galaxies, performed by Huchtmeier et al. (2000).

For excited galaxies, stellar masses, estimated via $L_K$, and hydrogen masses are characterized by the mean values $\langle\log M^*\rangle=7.52\pm0.20$ and $\langle\log M_{HI}\rangle=7.62\pm0.16$ in Solar masses. For quiescent galaxies, the corresponding mean values are $\langle\log M^*\rangle=6.96\pm0.15$ and $\langle\log M_{HI}\rangle=6.93\pm0.16$. The differences between two subsamples in the stellar mass,
$(0.56\pm0.25)$ dex, and in the hydrogen mass, $(0.69\pm0.23)$ dex, seem to be significant at a level of $(2-3) \sigma$. A part of this difference can be due to the different selection conditions for the subsamples. So, the excited galaxies are seen in a larger volume with the average distance of $(6.74\pm0.48)$ Mpc, while the dim quenched galaxies have a  shorter average distance of $(5.25\pm0.56)$ Mpc. Despite the difference in stellar masses of the two subsamples, both categories of galaxies have a gas-rich component. Taking into account a factor of 1.4, which accounts for the helium abundance and heavier elements, it can be asserted that more than half of the baryonic mass of these galaxies is still in a gaseous state. 
\begin{figure} 
\includegraphics[height=0.35\textwidth]{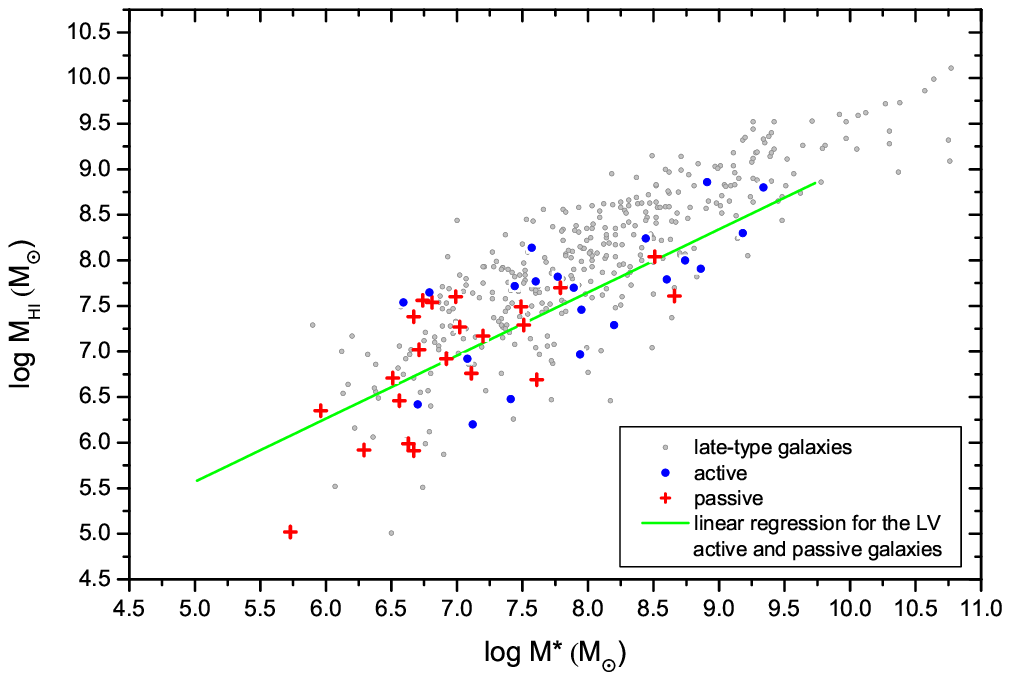}\\
\includegraphics[height=0.35\textwidth]{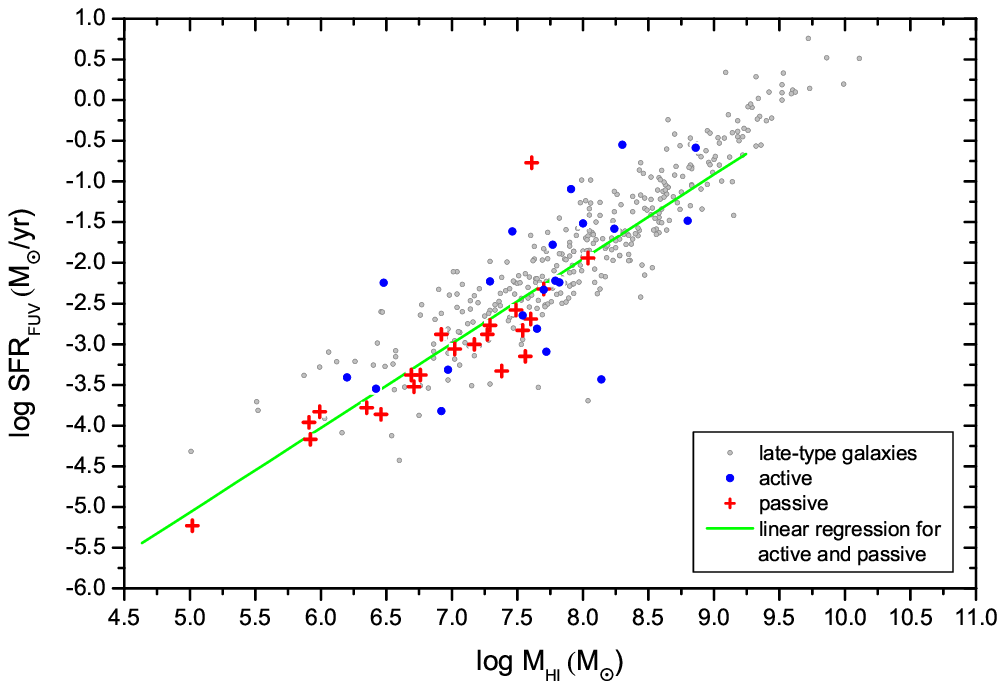}
\caption{The top panel: the relation between the hydrogen mass and stellar mass for excited galaxies (the circles) and quiescent galaxies (the crosses). The bottom panel: the relation between the integral star-formation rate and hydrogen mass for excited and quiescent galaxies. The parameters of linear regressions are specified in Table 3. The late-type galaxies of the total LV sample are shown with gray dots.} 
\end{figure} 

The top panel of Fig.5 shows the distribution of excited galaxies (the solid circles) and quiescent galaxies (the crosses) on hydrogen and stellar masses. Both subsamples follow near the same dependence with the regression line
\begin{equation}
\log(M_{HI})=(0.69\pm0.09)\times \log(M^*)+2.1\pm0.7.
   \end{equation}
indicated with the green colour. The late-type galaxies of the total LV sample (gray dots) are located approximately in the same area. Their regression line is about $0.25 dex$ above the green line.
Consequently, the relative amount of gas in a dwarf galaxy has only a slight effect on its burst activity. This would not be a case corresponding to that, when a starburst leads to the irreversible evacuation of a significant portion of gas from the galaxy body. 

The bottom panel of Fig.5 shows the distribution of galaxies on their star-formation rate and hydrogen mass. Both excited and quiescent galaxies are located approximately along the same dependence 
 \begin{equation}
\log[SFR(FUV)]=(1.04\pm0.11)\log(M_{HI})-10.3\pm0.8,
 \end{equation}
shown with the green line, although some galaxies ( Cam A, N4656UV) exhibit significant deviations from the regression line. The total sample of late-type galaxies in the Local Volume (gray dots) follows nearly the same relation.
The mean ratio $\log[SFR(FUV)/M_{HI}]$ equals to --10.0~yr$^{-1}$, i.e., a typical excited or quiescent galaxy is able to maintain the observed star-formation rate over further 10 Gyr. Taking account of the presence of He and other elements increases the average time of gas exhaustion up to $\sim14$~Gyr which coincides with the age of the Universe. 

\section{SFR in different environment conditions} 
Every galaxy in the Local Volume catalog has been supplied with the ``tidal index''
 \begin{equation}
 \Theta_1 = \max [\log(L_i/D_i^3)] + C, \,\,\,\,   i = 1, 2,....N,
 \end{equation}
where $L_i$ is a luminosity of neighboring galaxy in the $K$-band, $D_i$ is its
separation from the considered galaxy; ranking the surrounding galaxies on
the magnitude of their tidal force, $F_i \sim L_i/D_i^3$, allows to find the
most significant so-called ``Main Disturber'', MD; the constant $C$ is chosen
so that the galaxy with $\Theta_1 = 0$ is located at the ``zero velocity sphere''
relative to its MD; consequently, the causaly related galaxies with a
positive $\Theta_1$ are referred to as the population of the groups. The present values 
of $\Theta_1$ are taken from the latest version 
of the UNGC database (http://www.sao.ru/lv/lvgdb) updated with new estimates 
of galaxy distances.

Based on the data from different catalogs (Makarov \& Karachentsev 2011, Kourkchi \& Tully 2017), about 55\% of galaxies in the Local universe are members of groups and clusters of different population. The UNGC Local Volume catalog contains 60\% of galaxies belonging to groups. In our sample of 28 temporarily excited and 23 temporarily quiescent galaxies, the relative number of group members with the tidal index $\Theta_1>0$ is about the same (31/51). The top panel of Fig.6 presents the distribution of galaxies on their specific star-formation rate $SFR(FUV)/M^*$ and parameter $\Theta_1$. As one can see from this diagram, excited and quiescent galaxies are well mixed with each other. The regression line $\log sSFR=(0.03\pm0.04)\times\Theta_1-10.1\pm0.1$ has insignificant slope indicating some weak effect of the interaction of a galaxy with its nearest neighbor on an increase in the star-formation rate in it. The average specific star-formation rate of galaxies from our sample is equal to $-10.1\pm0.1$ (yr$^{-1}$) which practically coincides with the parameter $\log H_0=-10.14$~(yr$^{-1}$). This agreement, also noticed by Kroupa et al. (2020), again indicates the smouldering character of star formation in late-type dwarf galaxies at the cosmic time-scale.  

\begin{table*}
 \caption{Parameters of linear regressions $Y=k\times X+c$. }
 \begin{tabular}{llcrr}\hline
  
     Y			          & X	        &N   &$k\pm\sigma_k$	  & $c\pm\sigma_c$\\
 \hline 
 $\log SFR(H\alpha)$ (excited)                  &$\log M^*$	&28	 &0.79$\pm$0.10	   &$-8.0\pm0.8$ \\
 $\log SFR(H\alpha)$ (quiescent)                &$\log M^*$	&23	 &1.06$\pm$0.16	   &$-12.2\pm1.1$ \\
 $\log SFR(H\alpha)$ (the LV late-type)         &$\log M^*$     &425	 &1.09$\pm$0.03	   &$-10.9\pm0.2$ \\
 $\log SFR(FUV)$ (excited \& quiescent)         &$\log M^*$	&51	 &0.94$\pm$0.07	   &$-9.6\pm0.5$ \\
 $\log SFR(FUV)$ (the LV late-type)             &$\log M^*$	&425	 &0.95$\pm$0.02	   &$-9.5\pm0.2$ \\
 $\log M(HI)$ (excited \& quiescent)            &$\log M^*$	&43	 &0.69$\pm$0.09	   &$2.1\pm0.7$ \\
 $\log M(HI)$ (the LV late-type )               &$\log M^*$	&396	 &0.75$\pm$0.02	   &$2.0\pm0.2$ \\
 $\log SFR(FUV)$ (excited \& quiescent)         &$\log M(HI)$   &43      &1.04$\pm$0.11    &$-10.3\pm0.8$ \\
 $\log SFR(FUV)$ (the LV late-type )            &$\log M(HI)$   &396     &1.06$\pm$0.02    &$-10.3\pm0.2$ \\
 $\log sSFR(FUV)$ (excited \& quiescent)        &$\Theta_1$	&51	 &0.03$\pm$0.04    &$-10.1\pm0.1$ \\
 $\log SFR(H\alpha)-\log SFR(FUV)$ (excited)    &$\Theta_1$	&28	 &0.03$\pm$0.02	   &$0.50\pm0.04$ \\
 $\log SFR(H\alpha)-\log SFR(FUV)$ (quiescent)  &$\Theta_1$	&23	 &$-$0.04$\pm$0.06 &$-1.62\pm0.07$\\
 \hline\end{tabular}
 \end{table*}

\begin{figure} 
\includegraphics[height=0.35\textwidth]{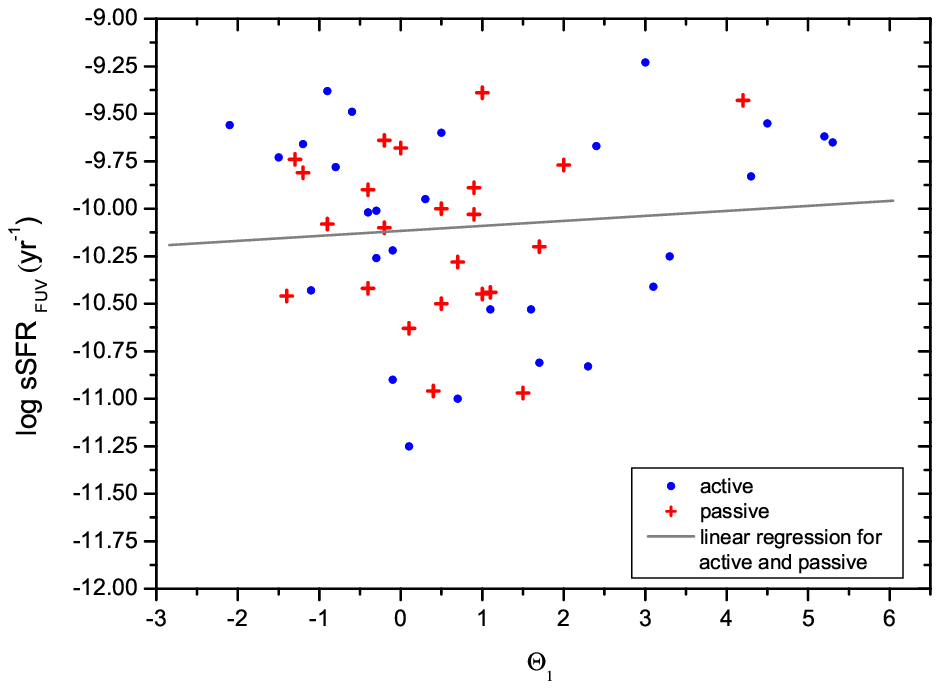}\\
\includegraphics[height=0.35\textwidth]{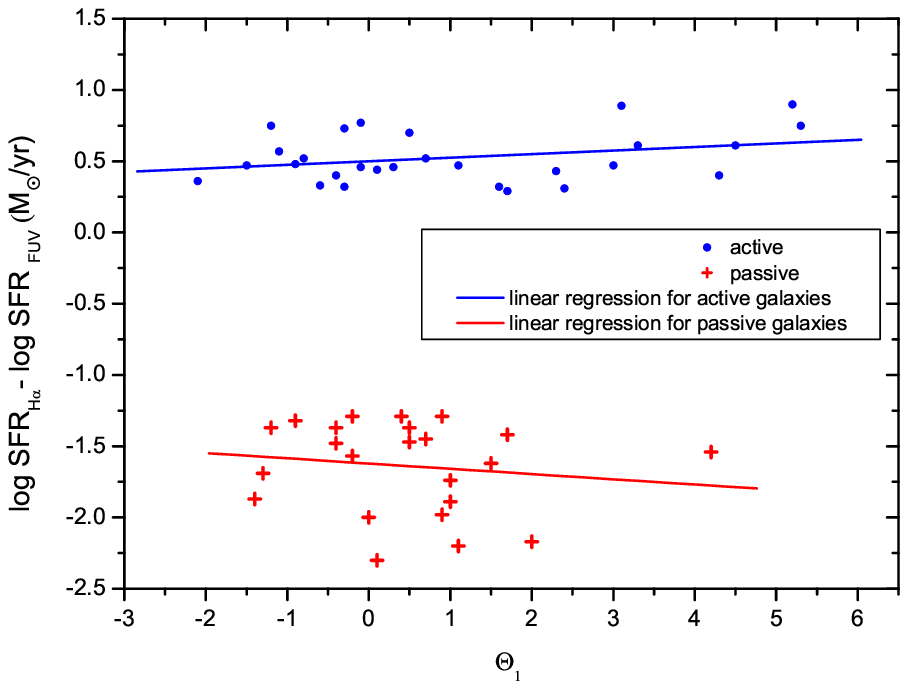}
\caption{The top panel: the specific star-formation rate vs. the tidal index $\Theta_1$ for excited (the circles) and quiescent (the crosses) galaxies. The bottom panel: the ratio of the star-formation rates on scales of $\sim10$~Myr and $\sim100$~Myr for excited and quiescent galaxies vs. the tidal index 
$\Theta_1$. Parameters of the linear regressions are specified in the last rows of Table 3.} \end{figure} 

The distribution of temporarily excited and temporarily quiescent galaxies on their ratio $\log[SFR(H\alpha)/SFR(FUV)]$ and the tidal index is shown in the bottom panel of Fig. 6. The last two rows in Table 3 show the linear regression parameters for excited and quiescent galaxies. The excited ones  reveal a slight tendency to the starburst activity growth with the presence of a close neighbor. The quiescent galaxies show a slight opposite tendency. In both cases, the slope of the regression line is comparable to the magnitude of the slope error. Consequently, the influence of a close neighbor on the star-formation activity in a galaxy is of secondary importance compared to the internal conditions that cause converting gas into stars.
Similar results were obtained by Hunter \& Elmegreen (2004), Lee et al. (2007), Bothwell et al. (2009), Lamastra et al. (2013), and Diaz-Garcia \& Knapen (2020) . This conclusion is also consistent with the results of the New Horizon cosmological simulations performed by Martin et al. (2021). 

\section{Discussion} Considering a sample of 477 late-type galaxies from the Local Volume with $SFR$ estimated via the H$\alpha$ and $FUV$ fluxes, we selected galaxies with the extreme values of the $SFR(H\alpha)/SFR(FUV)$ ratio (more than 2 and less than 1/20). Each subsample has about 5\% of the total number of galaxies. Almost all of them are dwarf galaxies with the stellar masses $\log(M^*/M_{\odot})<9.5$. ``Excited'' galaxies have stellar and hydrogen masses on average 4--5 times greater than those of ``quiescent'' galaxies, which is partly due to the difference in the detection conditions: fainter quiescent galaxies are seen mainly at shorter distances. Despite the huge difference in the $SFR(H\alpha)$ activity on the $\sim10$ Myr scale, galaxies of both categories show approximately the same specific star-formation rate $sSFR(FUV)$ on the scale of $\sim100$ Myr. In both temporarily excited and quiescent dwarf galaxies, the average specific star-formation rate is $\log[sSFR(FUV)]=-10.1\pm0.1$~(yr$^{-1}$) coinciding with the Hubble parameter $\log H_0=-10.14$~(yr$^{-1})$. Consequently, the gas-to-star conversion process in late-type dwarf galaxies has a steady smouldering character at the cosmic
time $H_0^{-1}$. At short time intervals, $t<100$~Myr, variations of $SFR$ have a typical amplitude of less than 1 order above the main sequence and a
fading amplitude of 1--2 order below the average level. 

Among excited dwarf galaxies, the average hydrogen mass, $\langle\log M_{HI}\rangle=7.62\pm0.16$, equals approximately to the average stellar mass,
$\langle\log M^*\rangle=7.52\pm0.20$. The same is valid also for quiescent dwarfs with their $\langle\log M_{HI}\rangle=6.93\pm0.16$ and
$\langle\log M^*\rangle=6.96\pm0.15$. Such an agreement would not have taken place if the starburst irreversibly pushed a significant portion of gas out of the potential well of a galaxy. Therefore, the late-type dwarf galaxies have gas reserves sufficient to maintain the observed star-formation rate over the next cosmic Hubble time. In the model of ``closed box'' evolution (i.e. without accretion of gas from the cosmic web and with gas recycling), the populations of Irr, Im, and BCD galaxies are near halfway through their gas-dynamical evolution. 

About 60\% of the LV galaxies are the members of groups of different multiplicity. The same proportion of group members occurs among galaxies with the extreme star-formation rate ratios. The presence of a gas-rich dwarf galaxy in the vicinity of a massive neighbor can either enhance or suppress star formation in it. Ranging the galaxies of our sample by the tidal force of the nearest neighbor, $\Theta_1$, shows only a slight excess of $SFR$ in excited galaxies and its absence in galaxies that are in the passive stage. The star formation in late-type dwarf galaxies is mainly driven by internal processes rather than external factors, which is consistent with the results obtained by Larson \& Tinsley (1978), Bergvall et al. (2003), Li et al. (2008), and Elison et al. (2013)

It should be recalled that the Local Volume contains about 350 known early-type dwarf galaxies (dSph and dE), which are gas-poor and with  depressed star formation. Most of these passive dwarfs are concentrated within the extended halos of massive neighbors. Some of them probably result from the evolution of irregular and BCD galaxies passing through the stage of transient-type dwarf (Tr). However, special features of $SFR$ of the population of dSph and dE galaxies remained beyond our consideration. 

{\bf Acknowledgments} 

{This work is supported by the RSF grant 19--12--00145. We thank the referee for many helpful suggestions, which improved the content, clarity and presentation of the paper.}

 {}

\clearpage
\section{Appendix A. Stellar masses and SFRs of late-type galaxies in the Local Volume.}
 
\begin{tabular}{lcrrr}\hline
 Galaxy name   &    RA (2000.0) DEC &  $\log M^*$ &$\log SFR_{H\alpha} $ &$\log SFR_{FUV}$ \\
\hline
               &                    &  $M_{\odot}$ & $M_{\odot}$/yr&  $M_{\odot}$/yr  \\
\hline    
UGC12894       &    000022.5+392944 &   7.35 &  -2.29   &   -2.03 \\
WLM            &    000158.1-152740 &   7.48 &  -2.68   &   -2.23 \\
ESO409-015     &    000531.8-280553 &   7.88 &  -1.47   &   -1.72 \\ 
AGC748778      &    000634.4+153039 &   6.17 &  -4.64   &   -3.53 \\
UGC00064       &    000744.0+405232 &   7.93 &  -1.51   &   -1.63  \\
ESO349-031     &    000813.3-343442 &   6.90 &  -4.03   &   -3.02 \\
NGC0024        &    000956.4-245748 &   9.26 &  -1.01   &   -0.60 \\
NGC0045        &    001403.9-231056 &   9.11 &  -0.54   &   -0.45 \\
NGC0055        &    001508.5-391313 &   9.26 &  -0.36   &   -0.22 \\
ESO410-005     &    001531.4-321048 &   6.67 & $<$-6.26   &   -3.96 \\
JKB129         &    002041.4+083701 &   7.17 &  -2.89   &   -2.54 \\
ESO294-010     &    002633.3-415120 &   6.07 &  -4.28   &   -3.81 \\
UGC00288       &    002904.0+432554 &   7.57 &  -2.61   &   -2.33 \\
ESO473-024     &    003122.5-224557 &   7.50 &  -2.10   &   -1.94 \\
And IV         &    004232.3+403419 &   7.00 &  -3.12   &   -2.42 \\
DDO226         &    004303.8-221501 &   7.48 &  -2.90   &   -2.46 \\
NGC0247        &    004708.3-204536 &   9.28 &  -0.55   &   -0.25 \\
NGC0253        &    004734.3-251732 &  10.76 &   0.26   &    0.34 \\
KDG002         &    004921.1-180428 &   6.63 & $<$-5.72   &   -3.83 \\
DDO006         &    004949.3-210058 &   6.86 &  -4.14   &   -2.86 \\
ESO540-032     &    005024.6-195425 &   6.61 &  -4.98   &   -3.91 \\
NGC0300        &    005453.5-374057 &   9.18 &  -0.78   &   -0.56 \\
LGS 3          &    010355.0+215306 &   5.73 & $<$-6.85   &   -5.23 \\
IC1613         &    010447.8+020800 &   7.88 &  -2.35   &   -2.04 \\
UGC00685       &    010722.3+164102 &   7.81 &  -2.25   &   -2.27 \\
\hline

\multicolumn{5}{l}{Note: The complete data table is available online.}
\end{tabular}

 \clearpage
\section{Appendix B. Brief description of some individual cases} 

{\em PGC~2448110.} The blue compact object on the northern periphery of the asymmetric Sm galaxy NGC~5474. Probably, it is the brightest HII region in this galaxy  with the absolute magnitude  $M_B=-11.9$ mag.

{\em Clump~III.} The intergalactic HII region in the M81 group with the coordinates J100040.4+683937 and absolute magnitude $M_B=-8.3$~mag. 

{\em JKB~83.} The intergalactic HII region in the M81 group with the coordinates J095549.6+691957 and $M_B=-8.4$~mag detected by James et al. (2017). 

{\em JKB~142.} The dwarf companion of the spiral galaxy NGC~628 with $M_B=-11.9$~mag and the coordinates J014548.2+162241 found by James et al. (2017). 

{\em N~2903-HI.} The dIrr companion of the spiral galaxy NGC~2903 with $M_B=-11.7$~mag and the coordinates J093039.9+214325 detected by Irwin et al. (2009)

{\em KDG~61em.} The compact HII region projected onto the dwarf spheroidal galaxy  KDG~61, the M81 companion. 

{\em Garland = PGC~029167.} The chain of HII regions projected onto the southern edge of NGC~3077 (Karachentsev et al. 1985). 

{\em NGC~2683dw1.} The dwarf companion of the spiral galaxy NGC~2683 with $M_B=-10.6$~mag and the coordinates J085326.8+331819, (Karachentsev et al. 2015b)

{\em LV~J0300+25 = AGES~J030039+254656.} The companion of the isolated galaxy NGC~1156 (Minchin et al. 2010). 

{\em KK~35.} The dIrr companion or a diffuse complex of HII regions on the periphery of the spiral galaxy IC~342. 

{\em N~4656UV.} The dIrr companion of the spiral galaxy NGC~4656 with $M_B=-16.6$~mag and the coordinates J124415.7+321700. Internal motions in this galaxy were studied by Zasov et al. (2017).

{\em KDG~215.} The companion of the spiral galaxy NGC~4826, the star-formation history of which has been studied by Cannon et al. (2018). 

\begin{thebibliography}{}

\bibitem{}Akins H.B., Christensen C.R., Brooks A.M., et al, 2021, ApJ, 909, 139
\bibitem{} Bergvall N., Laurikainen E., Aalto S., 2003, A\&A, 405, 31
\bibitem{}Bothwell M.S., Kennicutt R.C., Lee J.C., 2009, MNRAS, 400, 154
\bibitem{}Bouchard A., Da Costa G.S., Jerjen H., 2009, AJ, 137, 3038 
\bibitem{}Cannon J.M., Shen Z., McQuinn K.B.W., et al, 2018, ApJ, 864L, 14 
\bibitem{}Cote S., Draginda A., Skillman E.D., Miller B.W., 2009, AJ, 138, 1037 
\bibitem{} Diaz-Garcia S. \& Knapen J.H., 2020, A\&A, 635, A197
\bibitem{} Ellison S.L., Mendel J.T., Patton D.R., Shudder J.M., 2013, MNRAS, 435, 3627
\bibitem{}Flores Velazques J.A., Gurvich A.B., Faucher-Giguere C.F. et al. 2021, MNRAS, 501, 4812
\bibitem{}Gavazzi G., Boselli A., Arosio I., et al, 2006, A\&A, 446, 839 
\bibitem{}Gil de Paz A., Madore B.F., Pevunova O., 2003, ApJS, 147, 29 
\bibitem{}Grossi M., Disney M.J., Pritzl B.J., et al, 2007, MNRAS, 374, 107 
\bibitem{}Haynes M.P., Giovanelli R., Martin A.M., et al. 2011, AJ, 142, 170
\bibitem{}Huchtmeier W.K., Karachentsev I.D., Karachentseva V.E., Ehle M., 2000, A\&AS, 141, 469
\bibitem{}Hunter D.A., Elmegreen B.G., 2004, AJ, 128, 2170
\bibitem{}Irwin J.A., Hoffman G.L., Spekkens K., et.al, 2009, ApJ, 692, 1447 
\bibitem{}James B.L., Koposov S.E., Stark D.P., et al, 2017, MNRAS, 465, 3977 
\bibitem{}James P.A., Prescott M., Baldry I.K., 2008, A\&A, 484, 703 
\bibitem{}Kaisina E.I., Makarov D.I., Karachentsev I.D., Kaisin S.S., 2012, AstBull, 67, 115 
\bibitem{}Kaisin S.S., Karachentsev I.D., 2019, AstBu, 74, 1 
\bibitem{}Kaisin S.S., Karachentsev I.D., 2014, AstBu, 69, 390 
\bibitem{}Kaisin S.S., Karachentsev I.D., 2013, AstBu, 68, 381 
\bibitem{}Kaisin S.S., Karachentsev I.D., Ravindranath S., 2012, MNRAS, 425, 2083 
\bibitem{}Kaisin S.S., Karachentsev I.D., Kaisina E.I., 2011, Ap, 54, 315 
\bibitem{}Kaisin S.S., Karachentsev I.D., 2008, A \& A, 479, 603 
\bibitem{}Karachentseva V.E., Karachentsev I.D., Kashibadze O.G., 2020, Ap, 63, 151 
\bibitem{}Karachentsev I.D., Kaisina E.I., 2019, AstBu, 74, 111 
\bibitem{}Karachentsev I.D., Kaisin S.S., Kaisina E.I., 2015a, Ap, 58, 453 
\bibitem{}Karachentsev I.D., Sharina M.E., Makarov D.I., et.al, 2015b, Ap, 58, 309
\bibitem{}Karachentsev I.D., Kaisina E.I., 2013, AJ, 146, 46  
\bibitem{}Karachentsev I.D., Karachentseva V.E., Melnyk O.V., Courtois H.M., 2013a, AstBu, 68, 243
\bibitem{}Karachentsev I.D., Makarov D.I., Kaisina E.I., 2013b, AJ, 145, 101 (UNGC) 
\bibitem{}Karachentsev I.D., Kaisina E.I., Makarova L.N., 2011, MNRAS, 415L, 31 
\bibitem{}Karachentsev I.D., Kaisin S.S., 2010, AJ, 140, 1241 
\bibitem{}Karachentsev I.D., Kaisin S.S., 2007, AJ, 133, 1883 
\bibitem{}Karachentsev I.D., Karachentseva V.E., Boerngen F., 1985, MNRAS, 217, 731 
\bibitem{}Kennicutt R.C., Lee J.C., Funes S.J., et al, 2008, ApJS, 178, 247
\bibitem{}Koribalski B.S., Staveley-Smith L., Kilborn V.A. et al. 2004, AJ, 128, 16 
\bibitem{}Kourkchi E., Tully R.B., 2017, ApJ, 843, 16 
\bibitem{}Kovac K., Oosterloo T.A., van der Hulst J.M., 2009, MNRAS, 400, 743
\bibitem{}Kroupa P., Haslbauer M., Banik I., et al, 2020, MNRAS, 497, 37
\bibitem{}Lamastra A., Menci N., Fiore F., Santini P., 2013, A\&A, 552A, 44
\bibitem{}Larson R.B. \& Tinsley B.M., 1978, ApJ, 219, 46
\bibitem{}Lee J.C., Gil de Paz A., Tremonti C., et al, 2009, ApJ, 706, 599 
\bibitem{}Lee J.C., Kennicutt R.C., Funes S.J., et al, 2007, ApJL, 671, L113
\bibitem{}Lelli F., McGaugh S.S., Schombert J.M., 2016, AJ, 152, 157 
\bibitem{}Li C., Kauffmann G., Heckman T.M., et al. 2008, MNRAS, 385, 1903
\bibitem{}Makarov D.I., Karachentsev I.D., 2011, MNRAS, 412, 2498 
\bibitem{}Martin G., Jackson R.A., Kaviraj S., et al, 2021, MNRAS, 500, 4937 
\bibitem{}Martin D.C., Fanson J., Schiminovich D., et. al, 2005, ApJ, 619, L1 
\bibitem{}McQuinn K.B.W., Skillman E.D., Cannon J.M., et al, 2009, ApJ, 695, 561 
\bibitem{}Minchin R.F., Momjian E., Auld R., et.al, 2010, AJ, 140, 1093 
\bibitem{}Schiminovich D., Catinella B., Kauffmann G., et al. 2010, MNRAS, 408, 919
\bibitem{}Schlegel D.J., Finkbeiner D.P., Davis M., 1998, ApJ, 500, 525
\bibitem{}Sparre M., Hayward C.C., Springel V., et al. 2015, MNRAS, 447, 3548
\bibitem{}Stinson G.S., Dalcanton J.J., Quinn T., et al, 2007, ApJ, 667, 170 
\bibitem{}Verheijen M.A.W., 2001, ApJ, 563, 694
\bibitem{}Weisz D.R., Johnson B.D., Johnson L.C., 2012, ApJ, 744, 44 
\bibitem{}Zasov A.V., Saburova A.C., Egorov O.V. \& Uklein R.I., 2017, MNRAS, 469, 4370. 

\end{thebibliography}
\end{document}